\documentclass[12pt]{article}
\usepackage{epsfig,float}

\date{}
\begin{document}

\newcommand{\beq}{\begin{equation}}
\newcommand{\eeq}{\end{equation}}
\newcommand{\nn}{\nonumber}
\newcommand{\bea}{\begin{eqnarray}}
\newcommand{\eea}{\end{eqnarray}}

\title{Braneworlds and Dark Energy}

\author{Rui Neves\footnote{E-mail: rneves@ualg.pt}\\
{\small \it Centro Multidisciplinar de Astrof\'{\i}sica (CENTRA)}\\
{\small \it \& Faculdade de Ci\^encias e Tecnologia}\\{\small \it Universidade do
  Algarve}\\
{\small \it Campus de Gambelas, 8005-139 Faro, Portugal}\\
\mbox{}\\ 
Cenalo Vaz\footnote{Email: vaz@physics.uc.edu}\\ 
{\small \it Department of Physics, The University of Cincinnati}\\
{\small \it 400 Geology/Physics Building, PO Box 21001}\\{\small \it Cincinnati, OH 45221-0011, USA}
}

\maketitle

\begin{abstract}
In the Randall-Sundrum scenario, we analyse the dynamics of an $\mbox{AdS}_5$ 
braneworld when conformal matter fields propagate in five dimensions. We show 
that conformal fields of weight $-4$ are associated with stable
geometries which describe the dynamics of inhomogeneous dust,
generalized dark radiation and homogeneous polytropic dark energy on a spherically 
symmetric 3-brane embedded in the compact $\mbox{AdS}_5$ orbifold. We discuss aspects of the radion stability conditions and of the localization of gravity in 
the vicinity of the brane.

\end{abstract}

\section{Introduction}
In the Randall-Sundrum (RS) scenario \cite{RS1,RS2} 
the observable four-dimensional (4D) Universe is a 3-brane world embedded
in a $Z_2$ symmetric 5D anti-de Sitter (AdS) space. In the RS1 model
\cite{RS1} the fifth dimension is compact and there are two 3-brane
boundaries. The gravitational field is bound to the hidden positive
tension brane and decays towards the observable negative tension brane. 
In this setting, the hierarchy problem is reformulated as an
exponential hierarchy between the weak and Planck scales \cite{RS1}. 
In the RS2 model \cite{RS2}, the orbifold has an infinite fifth
dimension and just one observable positive tension brane near which
gravity is exponentially localized. 

In the RS models, the classical field dynamics is defined by 5D
Einstein equations with a negative bulk cosmological constant
$\Lambda_{\rm B}$, Dirac delta sources standing for the branes and a
stress-energy tensor describing other fields propagating in the bulk
\cite{RS1}-\cite{GW}. A set of vaccum solutions is given by ${\rm d}{\tilde{s}_5^2}=
{\rm d}{y^2}+{{\rm e}^{-2|y|/l}}{\rm d}{s_4^2}$, where $y$ is the
cartesian coordinate representing the fifth dimension, the 4D line
element ${\rm d}{s_4^2}$ is Ricci flat, $l$ is
the AdS radius given by $l=1/\sqrt{-{\Lambda_{\rm
          B}}{\kappa_5^2}/6}$ with
${\kappa_5^2}=8\pi/{{\rm M}_5^3}$ and ${\rm M}_5$ the fundamental 5D Planck
mass. In the RS1 model, the hidden Planck brane is located at $y=0$ and the 
visible brane at  $y'=\pi{r_{\rm c}}$, where
$r_{\rm c}$ is the RS compactification scale \cite{RS1}.
The brane tensions $\lambda>0$ and $\lambda'<0$ have the same absolute
value $|\lambda'|=\lambda$. In the vaccum $\lambda$ is given in terms of
$\Lambda_{\rm B}$ and 
$l$ by $\lambda=-{\Lambda_{\rm B}}l$. In the RS2 model
\cite{RS2}, the visible brane is the one with positive tension
$\lambda$ located at $y=0$. The hidden brane is sent to infinity and
is physically decoupled. 

The low energy theory of gravity on the observable brane is 
4D general relativity and the cosmology may be 
Friedmann-Robertson-Walker \cite{RS1}-\cite{TM}. 
In the RS1 model, this requires the stabilization of 
the radion mode using, for example, a 5D scalar
field \cite{GW,WFGK,CGRT,TM}. The gravitational 
collapse of matter has also been analyzed in the RS 
scenario \cite{CHR}-\cite{RC2}. However, an exact 5D geometry
describing a stable black hole localized on a 3-brane 
has not yet been discovered. Indeed, non-singular localized black holes have
only been found in an $\rm{AdS}_4$ braneworld \cite{EHM}. A solution to this
problem requires a simultaneous localization of gravity and matter which avoids
unphysical divergences \cite{CHR,KOP}-\cite{RC2} and
could be related to quantum black holes on the brane \cite{EFK}. In
addition, the covariant Gauss-Codazzi approach \cite{BC,SMS}
has uncovered a rich set of braneworld solutions, many of which have
not yet been associated with exact 5D spacetimes \cite{COSp}-\cite{RC1}.

In this paper we report on research about the dynamics of a
spherically symmetric 3-brane in the presence of 5D
conformal matter fields \cite{RC2,RC3} (see also \cite{EONO}). In the previous work
\cite{RC2,RC3} we have discovered a new class of exact
5D dynamical solutions for which gravity is bound to the brane by the 
exponential RS warp. These solutions were shown to be
associated with conformal bulk fields characterized by a stress-energy
tensor $\tilde{T}_\mu^\nu$ of weight -4 and by the equation of
state ${\tilde{T}_a^a}=2{\tilde{T}_5^5}$ (see also
\cite{KKOP} and \cite{IR}). They were also shown
to describe on the brane the dynamics of
inhomogeneous dust, generalized dark radiation and homogeneous
polytropic matter. However, the density and pressures
of the conformal bulk fluid increase with the coordinate of the fifth
dimension. Consequently and just like in the Schwarzschild black
string solution \cite{CHR}, the RS2 scenario is plagued with an
unphysical singularity at the AdS horizon. Such divergence does not
occur in the RS1 model because the compactified space ends before the 
AdS horizon is reached. However, the radion
mode turns out to be unstable \cite{RN}. In this
work we discuss new exact 5D braneworld solutions which are stable under
radion field perturbations and still describe on the visible brane the 
dynamics of inhomogeneous dust, generalized dark radiation and homogeneous
polytropic dark energy. We also consider the point of view of an effective
Gauss-Codazzi observer and show that the gravitational field is bound
to the vicinity of the brane.        

\section{5D Einstein equations and conformal fields}

To map the ${\rm AdS}_5$ orbifold, consider the coordinates 
$(t,r,\theta,\phi,z)$ where $z$ is related to the cartesian coordinate
$y$ by $z=l{{\rm e}^{y/l}}$, $y>0$. The most general non-factorizable
dynamical metric 
consistent with the $Z_2$ symmetry in $z$ and
with 4D spherical symmetry on the brane is given by 
\beq
{\rm d}{\tilde{s}_5^2}={\Omega^2}\left({\rm d}{z^2}-{{\rm e}^{2A}}{\rm
    d}{t^2}+{{\rm e}^{2B}}{\rm d}{r^2}+{R^2}{\rm d}{\Omega_2^2}\right),\label{gm1}
\eeq
where $\Omega=\Omega(t,r,z)$, $A=A(t,r,z)$, $B=B(t,r,z)$ and
$R=R(t,r,z)$ are $Z_2$ symmetric functions. $R(t,r,z)$ represents the
physical radius of the 2-spheres and $\Omega$ is the warp factor
characterizing a global conformal transformation on the metric.

In the RS1 model, the classical dynamics is defined by the 5D
Einstein equations,
\beq
{\tilde{G}_\mu^\nu}=-{\kappa_5^2}\left\{{\Lambda_{\rm B}}{\delta_\mu^\nu}+
{1\over{\sqrt{\tilde{g}_{55}}}}\left[\lambda\delta
\left(z-{z_0}\right)+\lambda'\delta\left(z-{{z'}_0}\right)\right]\left({\delta_\mu^\nu}-{\delta_5^\nu}
{\delta_\mu^5}\right)-{\tilde{\mathcal{T}}_\mu^\nu}\right\},
\label{5DEeq}
\eeq
where ${\tilde{\mathcal{T}}_\mu^\nu}$ is
the stress-energy tensor of the matter fields which is
conserved in 5D, 

\beq
{\tilde{\nabla}_\mu}{\tilde{\mathcal{T}}_\nu^\mu}=0\label{5Dceq}.
\eeq

For a general 5D metric $\tilde{g}_{\mu\nu}$, (\ref{5DEeq}) and
(\ref{5Dceq}) form an extremely complex system of differential equations. 
To solve it we need simplifying assumptions about
the field varia\-bles involved in the problem. Let us first consider that the bulk matter is
described by conformal fields with weight
$s$. Under the conformal transformation
${\tilde{g}_{\mu\nu}}={\Omega^2}{g_{\mu\nu}}$, the stress-energy tensor
satisfies
${\tilde{\mathcal{T}}_\mu^\nu}={\Omega^{s+2}}{\mathcal{T}_\mu^\nu}$. 
Consequently, (\ref{5DEeq}) and (\ref{5Dceq}) may be rewritten as \cite{RC2} 
\bea
&&{G_\mu^\nu}=-6{\Omega^{-2}}\left({\nabla_\mu}\Omega\right){g^{\nu\rho}}
{\nabla_\rho}\Omega+
3{\Omega^{-1}}{g^{\nu\rho}}{\nabla_\rho}{\nabla_\mu}\Omega
-3{\Omega^{-1}}{\delta_\mu^\nu}{g^{\rho\sigma}}{\nabla_\rho}{\nabla_\sigma}
\Omega\nonumber\\
&&-{\kappa_5^2}
{\Omega^2}\left\{{\Lambda_{\rm B}}{\delta_\mu^\nu}+{\Omega^{-1}}\left[\lambda
\delta(z-{z_0})+\lambda'\delta\left(z-{{z'}_0}\right)\right]\left(
{\delta_\mu^\nu}-{\delta_5^\nu}{\delta_\mu^5}\right)
-{\Omega^{s+2}}{\mathcal{T}_\mu^\nu}\right\},
\label{t5DEeq}
\eea
\beq
{\nabla_\mu}{\mathcal{T}_\nu^\mu}+{\Omega^{-1}}\left[(s+7){\mathcal{T}_\nu^\mu}{\partial_\mu}
\Omega-{\mathcal{T}_\mu^\mu}{\partial_\nu}\Omega\right]=0\label{t5Dceq}.
\eeq

If we separate the conformal tensor $\tilde{\mathcal{T}}_\mu^\nu$ in two sectors
$\tilde{T}_\mu^\nu$ and $\tilde{U}_\mu^\nu$ with the same weight $s$,
${\tilde{\mathcal{T}}_\mu^\nu}={\tilde{T}_\mu^\nu}+{\tilde{U}_\mu^\nu}$
where ${\tilde{T}_\mu^\nu}={\Omega^{s+2}}{T_\mu^\nu}$ and
${\tilde{U}_\mu^\nu}={\Omega^{s+2}}{U_\mu^\nu}$, and take $s=-4$ then it is possible to split (\ref{t5DEeq}) as follows
\beq
{G_\mu^\nu}={\kappa_5^2}{T_\mu^\nu},\label{r5DEeq}
\eeq
\bea
&6{\Omega^{-2}}{\nabla_\mu}\Omega{\nabla_\rho}
\Omega{g^{\rho\nu}}-
3{\Omega^{-1}}{\nabla_\mu}{\nabla_\rho}\Omega{g^{\rho\nu}}+3{\Omega^{-1}}
{\nabla_\rho}{\nabla_\sigma}\Omega{g^{\rho\sigma}}{\delta_\mu^\nu}=\nonumber\\
&-{\kappa_5^2}
{\Omega^2}\left\{{\Lambda_{\rm B}}{\delta_\mu^\nu}+{\Omega^{-1}}\left[\lambda
\delta(z-{z_0})+\lambda'\delta\left(z-{{z'}_0}\right)\right]\left(
{\delta_\mu^\nu}-{\delta_5^\nu}{\delta_\mu^5}\right)\right\}+{\kappa_5^2}{U_\mu^\nu}.\label{5DEeqwf}
\eea
On the other hand, the Bianchi identity implies
\beq
{\nabla_\mu}{T_\nu^\mu}=0.\label{5DceqT}
\eeq
Then (\ref{t5Dceq}) is in fact
\beq
{\nabla_\mu}{U_\nu^\mu}+{\Omega^{-1}}\left(3{\mathcal{T}_\nu^\mu}{\partial_\mu}
\Omega-{\mathcal{T}_\mu^\mu}{\partial_\nu}\Omega\right)=0.\label{u5Dceq}
\eeq
Note that (\ref{r5DEeq}) and (\ref{5DceqT}) are 5D
Einstein equations with conformal bulk fields, but without
a brane or bulk cosmological constant. They do not depend on the 
warp factor which is dynamically defined by (\ref{5DEeqwf}) and 
(\ref{u5Dceq}). The warp is then the only effect reflecting the
existence of the brane or of the bulk cosmological constant. We
emphasize that this is only possible for the special set of bulk fields
which have a stress-energy tensor with conformal weight $s=-4$.

Although the system of dynamical equations is now partially decoupled,
it remains difficult to solve. Note for instance that $\Omega$ depends
non-linearly on $A$, $B$ and $R$. In addition, it is affected by
$T_\mu^\nu$ and $U_\mu^\nu$. So consider the special setting
$A=A(t,r)$, $B=B(t,r)$, $R=R(t,r)$ and $\Omega=\Omega(z)$. Then (\ref{r5DEeq}) and (\ref{5DEeqwf}) lead to 
\beq
{G_a^b}={\kappa_5^2}{T_a^b},\label{4DECeq}
\eeq
\beq
{G_5^5}={\kappa_5^2}{T_5^5},\label{5DEeqz}
\eeq
\beq
6{\Omega^{-2}}{{({\partial_z}\Omega)}^2}+{\kappa_5^2}{\Omega^2}{\Lambda_{\rm
    B}}={\kappa_5^2}{U_5^5},\label{rswf1}
\eeq
\beq
\left\{3{\Omega^{-1}}{\partial_z^2}\Omega+{\kappa_5^2}{\Omega^2}
\left\{{\Lambda_{\rm B}}+{\Omega^{-1}}\left[\lambda\delta(z-{z_0})+\lambda'\delta(z-{{z'}_0})\right]\right\}\right\}{\delta_a^b}={\kappa_5^2}{U_a^b},\label{rswf2}
\eeq
where the latin indices represent the 4D coordinates
$t$, $r$, $\theta$ and $\phi$. Since according to (\ref{4DECeq}) and
(\ref{5DEeqz}) $T_\mu^\nu$ depends only on $t$ and $r$, (\ref{5DceqT}) becomes 
\beq
{\nabla_a}{T_b^a}=0.\label{5DceqT1}
\eeq
On the other hand, (\ref{rswf1}) and (\ref{rswf2}) imply that 
$U_\mu^\nu$ must be diagonal,
${U_\mu^\nu}=diag(-\bar{\rho},{\bar{p}_{\rm r}}$,\\${\bar{p}_{\rm
      T}},{\bar{p}_{\rm T}},{\bar{p}_5})$,
with the density $\bar{\rho}$ and pressures $\bar{p}_{\rm r}$,
$\bar{p}_{\rm T}$ satisfying 
$\bar{\rho}=-{\bar{p}_{\rm r}}=-{\bar{p}_{\rm T}}$. In addition,
$U_\mu^\nu$ must only depend on $z$. Consequently,
${\nabla_a}{U_b^a}=0$ is an identity. Then using (\ref{u5Dceq}) and
noting that ${T_\mu^\nu}={T_\mu^\nu}(t,r)$, we find
\beq
{\partial_z}{\bar{p}_5}+{\Omega^{-1}}{\partial_z}\Omega\left(2{U_5^5}-{U_a^a}\right)=0,\quad
2{T_5^5}={T_a^a}.\label{u5Dceq1}
\eeq
If ${U_\mu^\nu}(z)$ is a conserved tensor field like $T_\mu^\nu$, then
$\bar{p}_5$ must be constant. So (\ref{u5Dceq1}) leads to the
following equations of state: 
\beq
2{T_5^5}={T_a^a},\quad 2{U_5^5}={U_a^a}.\label{eqst2}
\eeq
Then we obtain ${\bar{p}_5}=-2\bar{\rho}$. $U_\mu^\nu$ is thus constant. On the other hand if ${T_\mu^\nu}=diag\left(-\rho,{p_{\rm r}},{p_{\rm T}},{p_{\rm T}},{p_5}\right)$ where
$\rho$, $p_{\rm r}$, $p_{\rm T}$ and $p_5$ are, respectively, the density and
pressures then its equation of state is rewritten as
\beq
\rho-{p_{\rm r}}-2{p_{\rm T}}+2{p_5}=0.\label{eqst3}
\eeq

Note that $\rho$, $p_{\rm r}$, $p_{\rm T}$ and $p_5$ must be 
independent of $z$, but may be functions of $t$ and $r$. The bulk matter is, however,
inhomogeneously distributed along the fifth dimension because
the physical energy density, $\tilde{\rho}(t,r,z)$, and pressures,
$\tilde{p}(t,r,z)$, are related
to $\rho(t,r)$ and $p(t,r)$ by the scale factor $\Omega^{-2}(z)$. Also
note that $T_\mu^\nu$ determines the dynamics on the
branes and that in the RS1 model, the two branes have identical cosmological
evolutions. On the other hand, it is also important to note that the
warp factor depends on the
conformal bulk fields only through $U_\mu^\nu$. Consequently, the role of
$U_\mu^\nu$ is to influence how
the gravitational field is warped around the
branes. In our previous work $U_\mu^\nu$
was set to zero \cite{RC2,RC3,RN}. The corresponding
braneworld solutions were warped by the exponential RS scale factor
and turned out to be
unstable under radion field perturbations \cite{RN}. So we also introduce $U_\mu^\nu$ as a stabilizing sector.

\section{Exact 5D warped solutions}

\begin{figure}[H]
   \center{\psfig{file=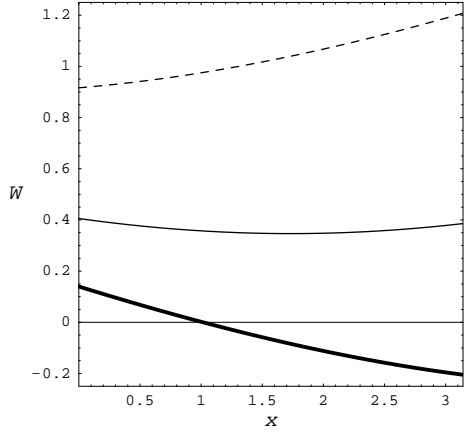,width=0.4\hsize}}
   \vspace{-0.2cm}
   \caption{Plots of $W=\ln\Omega(x)$, $x=y/{r_{\rm c}}$ for
     $l/{r_{\rm c}}=5$. The dashed, thin and thick 
     lines correspond, respectively, to
     $p_{\rm B}^5$ equal to $1.5,0.5$ and $0.15$.}
   \label{fig1:wfs}
 \end{figure}

The $\rm{AdS}_5$ braneworld dynamics is defined by the solutions of 
(\ref{4DECeq}) to (\ref{5DceqT1}) and (\ref{eqst3}). Let us first solve (\ref{rswf1}) and
(\ref{rswf2}). As we have seen, $U_\mu^\nu$ is constant with
$\bar{\rho}=-{\bar{p}_{\rm r}}=-{\bar{p}_{\rm T}}=-{\bar{p}_5}/2$. If
${\bar{p}_5}=0$ then ${U_\mu^\nu}=0$, and we end up with the
usual RS warp equations. As is well known, a solution is the
exponential RS warp $\Omega(y)={\Omega_{\rm
    RS}}(y)={{\rm e}^{-|y|/l}}$ \cite{RS1,RS2}. If $\bar{p}_5$ is non-zero then we find a new set of warp solutions. 
Integrating (\ref{rswf1}) and taking into
account the $Z_2$ symmetry, we obtain (see figure~\ref{fig1:wfs})
\beq
\Omega(y)={{\rm e}^{-|y|/l}}\left(1+{p_{\rm B}^5}{{\rm e}^{2|y|/l}}\right),\label{wfp5y}
\eeq
where ${p_{\rm B}^5}={\bar{p}_5}/(4{\Lambda_{\rm B}})$. This set of
solutions must also satisfy (\ref{rswf2}) which contains the 
Israel jump conditions. As a consequence, the brane tensions
$\lambda$ and $\lambda'$ are given by
\beq
\lambda={\lambda_{\rm RS}}{{1-{p_{\rm B}^5}}\over{1+{p_{\rm
        B}^5}}},\quad \lambda'=-{\lambda_{\rm RS}}{{1-{p_{\rm
        B}^5}\exp(2\pi{r_{\rm c}}/l)}\over{1+{p_{\rm
        B}^5}\exp(2\pi{r_{\rm c}}/l)}},\label{wft4}
\eeq
where ${\lambda_{\rm RS}}=6/(l{\kappa_5^2})$.
Note that in the limit ${p_{\rm B}^5}\to 0$, we obtain the RS warp and
also the corresponding brane tensions.

To determine the dynamics on the brane we need to solve (\ref{4DECeq}) and
(\ref{5DEeqz}) when $T_\mu^\nu$ 
satisfies (\ref{5DceqT1}) and (\ref{eqst3}). Note that as long as
$p_5$ balances $\rho, {p_{\rm
    r}}$ and $p_{\rm T}$ according to (\ref{5DEeqz}) and (\ref{eqst3}), the 4D equation of state is not constrained. Three
examples corresponding to inhomogeneous dust, generalized dark
radiation and homogeneous polytropic matter were analised in
\cite{RC2} and \cite{RC3}. 
\noindent The latter describes the
dynamics on the brane of dark energy in the form of a polytropic
fluid. The diagonal conformal matter may be defined by
\beq
\rho={\rho_{\rm P}},\;
{p_{\rm r}}+\eta{{\rho_{\rm P}}^\alpha}=0,\;{p_{\rm T}}={p_{\rm r}},\;{p_5}=-{1\over{2}}
\left({\rho_{\rm P}}
+3\eta{{\rho_{\rm P}}^\alpha}\right),
\label{dmeqst}
\eeq
where $\rho_{\rm P}$ is the polytropic energy
density and the parameters ($\alpha$, $\eta$) characterize
different polytropic phases.

\begin{figure}[H]
   \center{\psfig{file=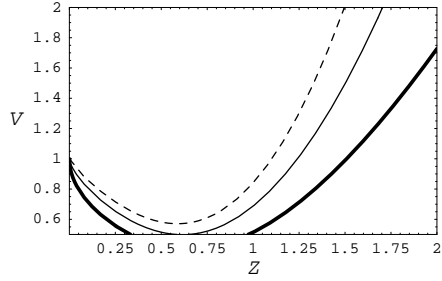,width=0.4\hsize}}
   \vspace{-0.2cm}
   \caption{Plots of $V=S{\dot{S}^2},\;Z={S^{1-\alpha}}$ for 
    $k>0$, $\eta>0$ and $a>0$. 
    The dashed, thin and thick lines correspond, respectively, to 
    $\alpha$ equal to $-1/4,-1/2$ and $-1$.}
   \label{fig2:l0ppp}
   \end{figure}

Solving the conservation equations, we find \cite{RC3,BBS}
\beq
{\rho_{\rm P}}={{\left(\eta+{a\over{S^{3-3\alpha}}}\right)}^
{1\over{1-\alpha}}},\label{dmdena}
\eeq
where $\alpha\not=1$, $a$ is an integration constant and $S=S(t)$ is the
Robertson-Walker scale factor of the brane world which is related to
the physical radius by $R=rS$. For $-1\leq\alpha<0$, the fluid is in
its generalized Chaplygin phase (see also \cite{BBS}). With this 
density, the Einstein equations lead to the following 5D dark energy polytropic solutions \cite{RC3}:
\beq
{\rm d}{\tilde{s}_5^2}={\Omega^2}\left[-{\rm d}{t^2}+{S^2}
\left({{{\rm d}{r^2}}\over{1-k{r^2}}}+{r^2}{\rm
    d}{\Omega_2^2}\right)\right]+{\rm d}{y^2},
\label{dmsol1}
\eeq
where the brane scale factor $S$ satisfies ${\dot{S}^2}={\kappa_5^2}
{\rho_{\rm P}}{S^2}/3-k$. The global evolution of the observable
universe is then given by \cite{RC1,RC3}
\beq
S{\dot{S}^2}=V(S)={{\kappa_5^2}\over{3}}{{\left(\eta{S^{3-3\alpha}}+
a\right)}^{1\over{1-\alpha}}}-k S.\label{dmdineq}
\eeq
In figure~\ref{fig2:l0ppp}, we present some ilustrative examples. 

\section{Radion stability}

To analyse how these solutions behave under radion field
perturbations, we consider the saddle point expansion of the RS action
\cite{RN,HKP,CGHW}
\beq
\tilde{S}=\int{{\rm d}^4}x{\rm d}y\sqrt{-\tilde{g}}
\left\{{\tilde{R}\over{2{\kappa_5^2}}}-
{\Lambda_{\rm B}}-{1\over{\sqrt{\tilde{g}_{55}}}}\left[\lambda\delta\left(y\right)+\lambda'\delta\left(y-\pi
  {r_{\rm c}}\right)\right]+
{\tilde{L}_{\rm B}}\right\},\label{5Dact1}
\eeq
where $\tilde{L}_{\rm B}$ is the lagrangian characterizing the 5D
matter fields. The most general metric
consistent with the 
$Z_2$ symmetry in $y$ and with 4D spherical symmetry on the
brane may be written in the form
\beq
{\rm d}{\tilde{s}^2}={a^2}{\rm d}{s_4^2}+{b^2}{\rm d}{y^2},\quad {\rm
  d}{s_4^2}=-{\rm d}{t^2}
+{{\rm e}^{2B}}{\rm d}{r^2}+{R^2}{\rm d}{\Omega_2^2},\label{5Dmt}
\eeq
where the metric functions $a=a(t,r,y)$, $B=B(t,r,y)$,
$R=R(t,r,y)$ and $b=b(t,r,y)$ are $Z_2$ symmetric. Now $a$ is the warp
factor and $b$ is related to the radion field. Our braneworld
backgrounds correspond to $b=1$, $B=B(t,r)$, $R=R(t,r)$ and $a=\Omega(y)$.

Consider (\ref{5Dmt}) with $a(t,r,y)=\Omega(y){{\rm e}^{-\beta(t,r)}}$ and
$b(t,r)={{\rm e}^{\beta(t,r)}}$. Then the dimensional reduction of
(\ref{5Dact1}) in the Einstein frame leads to \cite{RN}
\beq
\tilde{S}=\int{{\rm d}^4}x\sqrt{-{g_4}}\left({{R_4}\over{2{\kappa_4^2}}}-{1\over{2}}{\nabla_c}\gamma{\nabla_d}\gamma{g_4^{cd}}-\tilde{V}\right),\label{DR5Dact}
\eeq
where $\gamma=\beta/({\kappa_4}\sqrt{2/3}\,)$ is the canonically
normalized radion field. The function $\tilde{V}=\tilde{V}(\gamma)$ is
the radion potential, and it is given by
\begin{eqnarray}
&\tilde{V}={2\over{\kappa_5^2}}{\chi^3}\int{\rm d}y{\Omega^2}\left[3
{{({\partial_y}\Omega)}^2}+2\Omega
{\partial_y^2}\Omega\right]+\chi\int{\rm d}
y{\Omega^4}\left({\Lambda_{\rm B}}-{\tilde{L}_{\rm B}}\right)\nonumber\\
&+{\chi^2}\int{\rm d}y{\Omega^4}\left[\lambda\delta\left(y\right)+\lambda'\delta\left(y-\pi
  {r_{\rm c}}\right)\right],\label{rp}
\end{eqnarray}
where $\chi=\exp(-{\kappa_4}\gamma\sqrt{2/3}\,)$ and 
we have chosen ${\int_{-\pi{r_{\rm c}}}^{\pi{r_{\rm c}}}}{\rm d} y{\Omega^2}={\kappa_5^2}/{\kappa_4^2}$.

To analyse the stability of the $\rm{AdS}_5$ braneworld solutions, we
consider the saddle point expansion of the radion field potential
$\tilde{V}$. If ${p_{\rm B}^5}=0$, then $\Omega={\Omega_{\rm{RS}}}$. The radion
 potential has two critical extrema, ${\chi_1}=1$ and ${\chi_2}=1/3$ \cite{RN}. Our solutions correspond
to the first root ${\chi_1}=1$. The same happens if the bulk matter
is absent as in the RS vacuum solutions. Stable background solutions must be associated with a positive second variation of the radion potential. 
If the equation of state of the conformal bulk fields is independent
of the radion perturbation, then for $\chi={\chi_1}=1$ 
the second variation is negative, and so the braneworld solutions
are unstable \cite{RN}. 

\begin{figure}[H]
   \center{\psfig{file=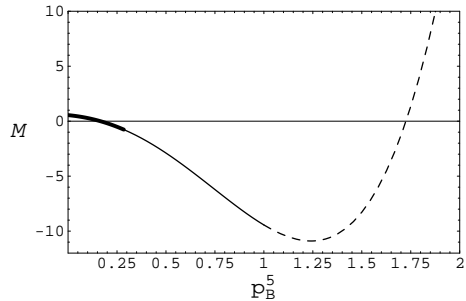,width=0.42\hsize}}
   \vspace{-0.2cm}
   \caption{Plot of radion mass parameter $M$ for
   $l/{r_{\rm c}}=5$. Thick line, $0<{p_{\rm B}^5}
     \leq{{\rm e}^{-2\pi/5}}: \lambda>0, \lambda'\leq 0$. Thin line,
     ${{\rm e}^{-2\pi/5}}
     <{p_{\rm B}^5}\leq 1: \lambda\geq 0,\lambda'>0$. Dashed line,
     ${p_{\rm B}^5}>1: 
     \lambda<0, \lambda'>0$.}
   \label{fig3:sta123}
   \end{figure}

If the equation of state is kept invariant
under the radion perturbations, it is possible to find stable solutions
at $\chi=1$ if the warp is changed. Indeed, the new
relevant warp functions are given in (\ref{wfp5y}). Consider 
$\tilde{\mathcal{V}}=\int{{\rm d}^4}x\sqrt{-{g_4}}\tilde{V}$. With
$x=y/{r_{\rm c}}$ and ${r_{\rm c}}{\int_{-\pi}^\pi}
     {\rm d}x{\Omega^2}={\kappa_5^2}/{\kappa_4^2}$, we find
\beq
{{{\delta^2}\tilde{\mathcal{V}}}\over{\delta{\gamma^2}}}
{{\Big |}_{\gamma=0}}=-{4\over{3}}{\kappa_4^2}{{\left({r_{\rm c}^2}\int
     {\rm d}x{\Omega^2}\right)}^{-1}}\int{{\rm d}^4}x\sqrt{-{\tilde{g}_4}}M,
\eeq 
where the dimensionless radion mass parameter $M$ is
\beq
M=\lambda{r_{\rm c}}{\kappa_5^2}{\Omega^4}(0)+\lambda'{r_{\rm c}}{\kappa_5^2}
      {\Omega^4}(\pi)-{{6{r_{\rm c}^2}}\over{l^2}}\int{\rm d}x{\Omega^4}.
\eeq
Stable solutions correspond to $M<0$. Consequently, stability exists
for a range of the model parameters if ${p_{\rm B}^5}>0$ (see
figure~\ref{fig3:sta123}). For ${p_{\rm B}^5}\leq 0$, all solutions are unstable.
\noindent For ${p_{\rm B}^5}>0$, the stability of the $\rm{AdS}_5$ braneworlds
also depends on the dimensionless ratio $l/{r_{\rm c}}$. For $l/{r_{\rm
         c}}<1.589\cdots$, all solutions turn out to be
     unstable. Stable universes begin to appear at $l/{r_{\rm
         c}}=1.589\cdots, {p_{\rm B}^5}=0.138\cdots$. For $l/{r_{\rm
         c}}>1.589\cdots$, we find stable solutions for an interval of 
$p_{\rm B}^5$ (see in figure \ref{fig3:sta123} the example of $l/{r_{\rm
         c}}=5$) which increases with $l/{r_{\rm c}}$. For large
         enough but finite $l/{r_{\rm c}}$, the stability interval
         approaches the limit $\left]0.267\cdots,3.731\cdots\right[$. Naturally, $M\to 0$ if $l/{r_{\rm c}}\to\infty$. 

\section{Gauss-Codazzi equations and localization of\\ gravity}

For an observer confined to the brane, 
the effective 4D Einstein equations are given by \cite{RC2,BC,SMS,RM}
\begin{eqnarray}
&{\mathcal{G}_\mu^\nu}={{2{\kappa_5^2}}\over{3}}
     \left[{\mathcal{U}_\alpha^\beta}
     {q_\mu^\alpha}{q_\beta^\nu}+\left({\mathcal{U}_\alpha^\beta}{n^\alpha}
     {n_\beta}-{1\over{4}}{\mathcal{U}_\alpha^\alpha}\right){q_\mu^\nu}
     \right]+{\mathcal{K}_\alpha^\alpha}{\mathcal{K}_\mu^\nu}\nonumber\\
&-{\mathcal{K}_\mu^\alpha}{\mathcal{K}_\alpha^\nu}-{1\over{2}}{q_\mu^\nu}
     \left({\mathcal{K}^2}-{\mathcal{K}_\alpha^\beta}{\mathcal{K}_\beta^\alpha}
     \right)-{\mathcal{E}_\mu^\nu},
\end{eqnarray}
where
${\mathcal{G}_\mu^\nu}={G_\alpha^\beta}{q^\alpha_\mu}{q_\beta^\nu}$, ${n^\mu}={\delta_5^\mu}$ is the unit
    normal to the brane and
    ${q_\mu^\nu}={\delta_\mu^\nu}-{n_\mu}{n^\nu}$. 
The stress-energy tensor is
${\mathcal{U}_\mu^\nu}=-{\Lambda_{\rm B}}{\Omega^2}(0){\delta_\mu^\nu}+
     {\mathcal{T}_\mu^\nu}$, ${\mathcal{K}_\mu^\nu}$ is the extrinsic
     curvature and ${\mathcal{E}_\mu^\nu}$ the traceless
     projection of the 5D Weyl tensor. The 4D observer finds the same
     dynamics on the brane because \cite{RC2}
\beq
{\mathcal{E}_a^b}={{\kappa_5^2}\over{3}}
     \left(-{T_a^b}+{1\over{2}}{T_5^5}{\delta_a^b}
     \right)
\eeq
and 
\beq
{4\over{3}}
     \left({U_a^b}+{1\over{4}}{U_5^5}{\delta_a^b}\right)-\left({\Lambda_{\rm B}}+
     {{{\kappa_5^2}{\lambda^2}}\over{6}}\right){\delta_a^b}\Omega^2(0)=0.
\eeq
Since the tidal acceleration \cite{RC2,RM} is
${a_{\rm T}}={\kappa_5^2}{\Lambda_{\rm B}}{{(1+{p_{\rm B}^5})}^2}/6<0$,
the gravitational field is bound to the vicinity of the brane.  

\section{Conclusions}

In this paper we have analised exact
5D solutions describing the dynamics of $\mbox{AdS}_5$ 
braneworlds when conformal fields of weight -4 are present in the bulk. 
We have discussed their behaviour under
radion field perturbations and shown
that if the equation of state characterizing the conformal fluid is
independent of the perturbation, then the radion may be stabilized by a sector of the conformal fields while
    another sector of the same class of fields generates the dynamics
    on the brane. Stabilization requires a bulk fluid sector with a constant
    negative 5D pressure and involves new warp functions. On the brane
    these solutions are able to describe, for example, the dynamics of
inhomogeneous dust, generalized dark radiation and homogeneous
polytropic dark energy. More general 4D equations of state may also be
considered. This analysis is left for future work. We have also shown
that an effective Gauss-Codazzi
observer sees gravity localized near the brane and deduces
the same dynamics on the brane if she makes the same hyphotesis about
the 5D fields. Whether gravity is suficiently bound to the brane and the
hierarchy strong enough are open problems for future research.
\vspace{1cm}

\leftline{\large \bf Acknowledgements}
\vspace{0.25cm}

We would like to thank the financial support of Funda\c {c}\~ao
  para a Ci\^encia e a Tecnologia (FCT) and Fundo Social
  Europeu (FSE) under the contract
SFRH/BPD/7182\\/2001 (III Quadro Comunit\'ario de Apoio), of Centro
  Multidisciplinar de Astrof\'{\i}sica (CENTRA) with project
  FJ01-CENTRA and Conselho de Reitores das Universidades
  Portuguesas (CRUP) with project Ac\c {c}\~ao Integrada
  Luso-Espanhola E-126/04.


\begin{thebibliography}{40}

\bibitem{RS1}
Randall L and Sundrum R 1999 {\it Phys. Rev. Lett.} {\bf 83} 3370

\bibitem{RS2}
Randall L and Sundrum R 1999 {\it Phys. Rev. Lett.} {\bf 83} 4690

\bibitem{GW}
Goldberger W D and Wise M B 1999 {\it Phys. Rev. D} {\bf 60} 107505

Goldberger W D and Wise M B 1999 {\it Phys. Rev. Lett.} {\bf 83} 4922

Goldberger W D and Wise M B 2000 {\it Phys. Lett. B} {\bf 475} 275

\bibitem{NK}
Kaloper N 1999 {\it Phys. Rev. D} {\bf 60} 123506

Nihei T 1999 {\it Phys. Lett. B} {\bf 465} 81
 
Cs\'aki C, Graesser M, Kolda C and Terning J 1999 {\it Phys. Lett. B}
 {\bf 462} 34

Cline J M, Grojean C and Servant G 1999 {\it Phys. Rev. Lett.} {\bf 83} 4245

\bibitem{KKOP}
Kanti P, Kogan I I, Olive K A and Pospelov M 1999 {\it Phys. Lett. B} {\bf
  468} 31

Kanti P, Kogan I I, Olive K A and Pospelov M 2000 {\it Phys. Rev. D} {\bf 61} 106004

\bibitem{WFGK}
DeWolf O, Freedman D Z, Gubser S S and Karch A 2000 {\it Phys. Rev. D}
{\bf 62} 046008

\bibitem{GT}
Garriga J and Tanaka T 2000 {\it Phys. Rev. Lett.} {\bf 84} 2778

\bibitem{GKR}
Giddings S, Katz E and Randall L 2000 {\it J. High Energy Phys.} JHEP0003(2000)\newline023

\bibitem{CGRT}
Cs\'aki C, Graesser M, Randall L and Terning J {\it Phys. Rev. D} {\bf
  62} 045015

\bibitem{TM}
Tanaka T and Montes X 2000 {\it Nucl. Phys.} {\bf B582} 259

\bibitem{CHR}
Chamblin A, Hawking S W and Reall H S 2000 {\it Phys. Rev. D} {\bf 61} 065007

\bibitem{EHM}
Emparan R, Horowitz G T and Myers R C 2000 {\it J. High Energy Phys.} 
JHEP0001(2000)007

Emparan R, Horowitz G T and Myers R C 2000 {\it
  J. High Energy Phys.} JHEP0001(2000)021

\bibitem{KOP}
Kanti P, Olive K A and Pospelov M 2000 {\it Phys. Lett. B} {\bf 481}
386
 
Shiromizu T and Shibata M 2000 {\it Phys. Rev. D} {\bf 62} 127502
 
L\"u H and Pope C N 2001 {\it Nucl. Phys.} {\bf B598} 492

Chamblin A, Reall H R, Shinkai H a and Shiromizu T 2001 {\it Phys. Rev. D}
{\bf 63} 064015

\bibitem{KT}
Kanti P and Tamvakis K 2002 {\it Phys. Rev. D} {\bf 65} 084010 

Kanti P, Olasagasti I and Tamvakis K 2003 {\it Phys. Rev. D} {\bf 68} 124001 

\bibitem{EFK}
Emparan R, Fabbri A and Kaloper N 2002 {\it J. High Energy Phys.}0208(2002)043

Emparan R, Garcia-Bellido J and Kaloper N 2003 {\it J. High Energy Phys.}0301\newline(2003)079

\bibitem{RC2}
Neves R and Vaz C 2003 {\it Phys. Rev. D} {\bf 68} 024007

\bibitem{BC}
Carter B 1993 {\it Phys. Rev. D} {\bf 48} 4835

Capovilla R and Guven J 1995 {\it Phys. Rev. D} {\bf 51} 6736

Capovilla R and Guven J 1995 {\it Phys. Rev. D} {\bf 52} 1072

\bibitem{SMS}
Sasaki M, Shiromizu T and Maeda K I 2000 {\it Phys. Rev. D} {\bf 62}
024008

Shiromizu T, Maeda K I and Sasaki M 2000 {\it Phys. Rev. D} {\bf 62} 024012

\bibitem{COSp}
Garriga J and Sasaki M 2000 {\it Phys. Rev. D} {\bf 62} 043523

Maartens R, Wands D, Bassett B A and Heard I P C 2000 {\it
  Phys. Rev. D} {\bf 62} 041301
 
Kodama H, Ishibashi A and Seto O 2000 {\it Phys. Rev. D} {\bf 62}
064022
 
Langlois D 2000 {\it Phys. Rev. D} {\bf 62} 126012

van de Bruck C, Dorca M, 
Brandenberger R H and Lukas A 2000 {\it Phys. Rev. D} {\bf 62} 123515 

Koyama K and Soda J 2000 {\it Phys. Rev. D} {\bf 62} 123502 

Langlois D, Maartens R, Sasaki M and Wands D 2001 {\it Phys. Rev. D} {\bf 63} 084009

\bibitem{DMPR}
Dadhich N, Maartens R, Papadopoulos P and Rezania V 2000 {\it
  Phys. Lett. B} 
{\bf 487} 1

Dadhich N and Ghosh S G 2001 {\it Phys. Lett. B} {\bf 518} 1
 
Germani C and Maartens R 2001 {\it Phys. Rev. D} {\bf 64} 124010

Bruni M, Germani C and Maartens R 2001 {\it Phys. Rev. Lett.} {\bf 87} 231302 

Govender M and Dadhich N 2002 {\it Phys. Lett. B} {\bf 538} 233

\bibitem{RM}
Maartens R 2000 {\it Phys. Rev. D} {\bf 62} 084023

\bibitem{RC1}
Neves R and Vaz C 2002 {\it Phys. Rev. D} {\bf 66} 124002

\bibitem{RC3}
Neves R and Vaz C 2003 {\it Phys. Lett. B} {\bf 568} 153

\bibitem{EONO}
Elizalde E, Odintsov S D, Nojiri S and Ogushi S 2003 {\it Phys. Rev. D} {\bf
  67} 063515 

Nojiri S and Odintsov S D 2003 {\it
  J. Cosmol. Astropart. Phys.} JCAP0306(2003)004

\bibitem{IR}
Rothstein I Z 2001 {\it Phys. Rev. D} {\bf 64} 084024

\bibitem{RN}
Neves R 2004 {\it TSPU Vestnik } Natural
and Exact Sciences 7 94

\bibitem{BBS}
Bento M C, Bertolami O and Sen S S 2002 {\it Phys. Rev. D} {\bf 66}
043507

Bento M C, Bertolami O and Sen S S 2003 {\it Phys. Rev. D} {\bf 67}
063003
 
Bento M C, Bertolami O and Sen S S 2004 {\it Phys. Rev. D} {\bf 70}
083519
 
Bento M C, Bertolami O, Santos N M C and Sen S S 2005 {\it Phys. Rev. D} {\bf
  71} 063501 
 
\bibitem{HKP}
Hofmann R, Kanti P and Pospelov M 2001 {\it Phys. Rev. D} {\bf 63}
124020
 
Kanti P, Olive K A and Pospelov M 2002 {\it Phys. Lett. B} {\bf 538} 146

\bibitem{CGHW}
Carroll S M, Geddes J, Hoffman M B and Wald R M 2002 {\it Phys. Rev. D}
{\bf 66} 024036

\end{thebibliography}
\end{document}